\documentclass[prl,twocolumn,superscriptaddress,floats]{revtex4b4}

\usepackage[dvips]{graphicx}

\usepackage{amsmath}

\usepackage{dcolumn}

\usepackage{rotating}

\newlength{\La} 
\newlength{\Lb} 
\newlength{\Lc} 

\newcolumntype{d}{D{.}{.}{-1}}

\newcommand{\srruo}{Sr$_2$RuO$_4$}
\newcommand{\srto}{Sr$_2$Ru$_{1-x}$Ti$_x$O$_4$}
\newcommand{\srton}{Sr$_2$Ru$_{0.91}$Ti$_{0.09}$O$_4$}
\newcommand{\caruo}{Ca$_2$RuO$_4$}

\newcommand{\csruo}{Ca$_{2-x}$Sr$_{x}$RuO$_4$}

\begin{document}

\advance\vsize by 2 cm
\title{Incommensurate magnetic ordering in \srto }

\author{M.~ Braden}
\affiliation{Forschungszentrum Karlsruhe, IFP, Postfach 3640, D-76021
Karlsruhe, Germany}
\affiliation{Laboratoire L\'eon Brillouin,
C.E.A./C.N.R.S., F-91191 Gif-sur-Yvette CEDEX,
France}

\author{O. Friedt}
\affiliation{Laboratoire L\'eon Brillouin,
C.E.A./C.N.R.S., F-91191 Gif-sur-Yvette CEDEX,
France}

\author{Y. Sidis}
\affiliation{Laboratoire L\'eon Brillouin,
C.E.A./C.N.R.S., F-91191 Gif-sur-Yvette CEDEX,
France}

\author{P. Bourges}
\affiliation{Laboratoire L\'eon Brillouin,
C.E.A./C.N.R.S., F-91191 Gif-sur-Yvette CEDEX,
France}

\author{M. Minakata}
\affiliation{Department of Physics, Kyoto University, Kyoto 606-8502, Japan}

\author{Y. Maeno}
\affiliation{Department of Physics, Kyoto University, Kyoto 606-8502, Japan}
\affiliation{CREST, Japan Science and Technology Corporation, Japan}

\date{\today, \textbf{DRAFT}}

\pacs{PACS numbers:}

\begin{abstract}

In $\rm Sr_2RuO_4$ the spin excitation spectrum is dominated by
incommensurate fluctuations at 
{\bf q}=($0.3$ $0.3$ q$_z$), 
which arise from Fermi-surface nesting.
We show that upon Ti substitution , known to suppress superconductivity,
a short range magnetic order 
develops with  a propagation vector 
($0.307$ $0.307$ $1$).
This finding confirms that  superconducting $\rm Sr_2RuO_4$ is extremely close to 
an incommensurate spin density wave instability. 
In addition,
the ordered moment in \srton ~ points along the c-direction, which 
indicates that the incommensurate spin fluctuations
exhibit the anisotropy required to explain a p-wave spin triplet pairing.



\end{abstract}

\maketitle

\srruo ~ the only superconducting layered perovskite isostructural
with the cuprates
has attracted considerable interest \cite{maeno}.
Though there is some evidence that the pairing in this compound
is  unconventional and of triplet character, the coupling
mechanism is still subject of debate\cite{ishida,julien,sigrist-a,
zhitomirsky}.
Inspired by the ferromagnetism in the 3D-perovskite SrRuO$_3$,
it has been proposed that the coupling is mediated by ferromagnetic
fluctuations \cite{rice}, for which, however, 
no experimental evidence has been found so far.
Instead, band structure calculations reveal magnetic fluctuations at
incommensurate positions arising from Fermi-surface nesting in the
1D-like bands associated to the d$_{xz}$- and d$_{yz}$-orbitals
\cite{mazin}.
Inelastic neutron scattering has perfectly confirmed this
nesting scenario\cite{sidis} with peaks in the  
dynamic magnetic susceptibility appearing at
{\bf q}=(0.3 0.3 $q_z$). We use 
the notation {\bf Q}=$\boldsymbol{\tau}$+{\bf q}, with {\bf Q} the 
scattering vector, {\bf q} the propagation vector in the first Brillouin-zone 
and $\boldsymbol{\tau}$ a reciprocal lattice vector;
all vectors are given in 
reciprocal lattice units corresponding to 
(${ 2 \pi \over a}$,${ 2 \pi \over a}$,${ 2 \pi \over c}$).

The pairing in \srruo ~ could be rather complicated since the Fermi
surface is formed by three bands. Of the three bands
only two contribute directly to the dynamical nesting.
The third band, the $\gamma$-band
associated to the d$_{xy}$-orbitals, is 
thought to be closer to ferromagnetism
with a possibly stronger role for the superconducting pairing.
However, recent calculations still do not give evidence
for ferromagnetic fluctuations arising from the $\gamma$-band
\cite{sigrist-n}.
In addition there is no evidence for  a dominant role
of ferromagnetic fluctuations in the transport properties \cite{julien},
and the superconducting transition temperature decreases upon pressure
increase \cite{shirakawa}, 
though ferromagnetic fluctuations should be enhanced.
The question about the dominant magnetic interaction
is, hence, still of strong interest.

Substitution may allow to identify the character of the
magnetic interaction, when static
magnetic order is induced.
Complete Ca-substitution leads to an antiferromagnetic
insulator \cite{caruo}. 
The structural properties of \caruo ~ are, however,
quite distinct from those of \srruo ; in particular the in plane
RuO distances are significantly enhanced. Therefore, the antiferromagnetic
order in \caruo ~ is not necessarily relevant for the Sr compound.
The intermediate \csruo -compositions present strongly enhanced magnetic
susceptibility near x=0.5 \cite{casrruo}, 
again in a region far away from the
superconducting compound; nevertheless this observation indicates
some ferromagnetic instability also in the 214-system.

Replacing a part of the Ru  by 
non-magnetic four-valent Ti 
(configuration 3d$^0$) leads
to magnetic anomalies even for very 
low Ti-concentration, x$>$0.025
in \srto \cite{minakata}.
These anomalies have been interpreted in terms of an
appearance of weak magnetic moments around Ti impurities.
In this work we show that these Ti-doped samples exhibit incommensurate 
magnetic ordering
corresponding to the Fermi-surface nesting instability.
Concerning the pure compound, one may conclude that it is 
much closer to an incommensurate spin density wave instability
than to a ferromagnetic one.

Two single crystals of \srto ~ with x=0.025 and 0.09 and 
a volume of about 100mm$^3$ each
were grown with a floating zone method in an infrared image furnace
\cite{minakata}. 
The inclusion of the Ti was verified by electron probe microanalysis. 
Furthermore, there is a clear reduction of the c-lattice 
constant and a shift of the frequency of the rotational phonon mode
from 7.86\ meV in the  pure compound \cite{phon-pap} 
to 9.97\ meV in \srton .
This agrees to the smaller ionic radius of Ti compared to Ru which 
should shift \srton ~ further away from the rotational structural
instability.
Magnetic neutron scattering experiments were performed on the 
triple axis spectrometers 1T (thermal source) and 4F (cold source) 
at the Orph\'ee
reactor using pyrolithic graphite (PG) monochromators and analyzers.
Higher order contamination was suppressed by either PG or cooled 
Be-filters.

\begin{figure}[t]
\includegraphics*[width=0.95\columnwidth]{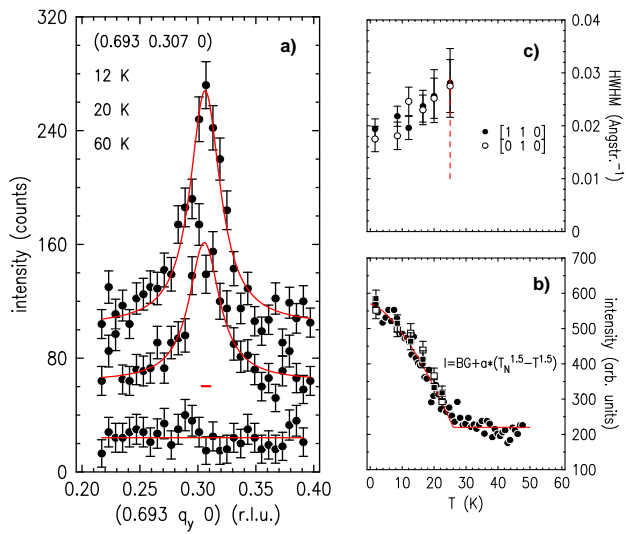}%
\caption{  Left part a) : 
scans across the incommensurate magnetic Bragg peaks at different
temperatures obtained on the 4F-spectrometer with $k_i$=1.48\AA$^{-1}$.
The y-axis scales of the scans at 20 and 60\ K are shifted for clarity
by 40 and 80 counts, respectively. The horizontal bar indicates the 
spectrometer resolution. 
Right part :
temperature dependence of the peak maximum intensity b) and the 
half width at half maximum of the Lorentzian profiles c)
observed at the incommensurate Bragg point (0.693 0.307 0).
In b) the circles present the raw intensity at the maximum of the peak, 
i.e. at (0.693 0.307 0), and the squares that arising from a fit to scans 
along the [110] (closed) and [010] (open squares) directions.
The line corresponds to a power-law fit.}
\label{fig:1}
\end{figure}

The search for commensurate magnetic order did not reveal
intensity neither at ferromagnetic nor at antiferromagnetic scattering vectors
for \srton .
The dominant features in the magnetic excitation 
spectrum in \srruo ~ being observed at incommensurate positions \cite{sidis} 
it appeared promising to look for 
corresponding order in \srton .
Indeed we found an elastic peak appearing very close to the 
positions of the inelastic scattering in \srruo . 
Due to the better peak to background ratio we performed an extensive
study on the cold triple axis 4F using long-wavelength neutrons,
($k_i$=1.48\AA$^{-1}$, $\lambda$=4.25\AA).
Figure 1 shows elastic scans across the incommensurate magnetic Bragg
point for different temperatures.
The elastic intensity appears below $T_N\sim$25\ K, however,
it is evident that the magnetic order remains limited in 
its spatial extension even at low temperature.
The maximum intensity was found at {\bf Q$_i$}=(0.693 0.307 0) which indicates
the propagation vector 
{\bf q$_i$}=(0.307 0.307 1), note that (1 0 0) is not 
a zone-center in the I4/mmm lattice.

The  second crystal with a lower Ti-content of x=0.025 does not
exhibit similar magnetic ordering. 
The comparison of the elastic scans between 20 and 1.6\ K indicates
a low-temperature intensity at {\bf Q$_i$} roughly a factor 20 less than in the x=0.09-sample
(relative to the intensity of a fundamental reflection) .
This finding agrees well with the appearance of
a sizeable ferromagnetic moment for concentrations higher than x=0.03
\cite{minakata}.

The intensity profiles measured for \srton ~
were fitted to Lorentzians folded with the 
experimental resolution. The resulting peak heights and  
resolution corrected half widths at
half maximum, $\kappa$, are shown in figure 1 as a function of temperature.
The peak height and the intensity at the peak maximum
indicate a sluggish phase transition near 25\ K.
In a weakly  antiferromagnetic metal the squared ordered moment
measured by the intensity of the superstructure Bragg peak
should be proportional to $(T_N^{1.5}-T^{1.5})$,
while the temperature dependence in a nesting-model may be more complex
\cite{moriya}. 
The fit by this power law is not very satisfactory, since the transition is 
definitely not sharp.
This behavior suggests a crossover from damped inelastic
fluctuations to elastic or quasi-elastic order at low temperature,
somehow reminiscent of a continuous slowing down observed in spin glass systems.
Furthermore, the spin-spin correlation length in the RuO$_2$-planes
remains finite and changes only gradually below the transition.
It saturates at $\sim$50\AA ~ at 1.6\ K. At this temperature, 
we do not find any evidence for a finite energy-width of the signal,
the intrinsic width may be estimated to be lower than 0.06\ meV.

\begin{figure}[t]
\includegraphics*[width=0.5\columnwidth]{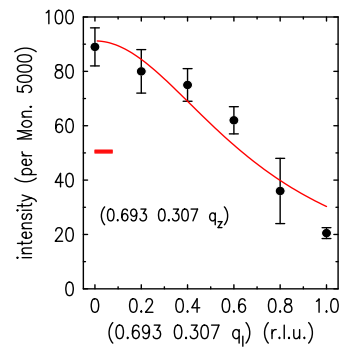}%
\caption{ Dependence of the peak height  of the Lorentzian profiles
at the incommensurate Bragg point (0.693 0.307 $q_z$)
on the out of plane component of the scattering vector.
The horizontal bar indicates the experimental resolution
along this direction and the solid line is a fit with a Lorentzian.
} \label{fig:2}
\end{figure}

The magnetic correlation is isotropic within the planes, but there is only
little modulation of the signal along $c$. 
Figure 2 shows the peak height of scans across 
(0.693 0.307 $q_z$) as a function of $q_z$. 
The $q_z$-dependence has been described by a Lorentzian yielding
a half width at half maximum of 0.7 reduced lattice units
or 0.35\AA$^{-1}$.
Such broad signals may not be interpreted in terms of 
a correlation length; nevertheless, the broadening indicates that the 
inter-plane coupling is restricted to the nearest neighbors.
The weak inter-plane coupling reflects the character of the incommensurate
fluctuations  \cite{sidis} which was reported to be essentially 
two-dimensional \cite{raymond}.
The intensity at the incommensurate peak maximum is rather 
weak compared to that in a fundamental nuclear reflection,
a few $10^{-5}$ for x=0.09;
however, due to the enormous broadening of the magnetic peaks the
ratio of the integrated intensities is much larger, close 
to $10^{-3}$.

In order to determine the incommensurate magnetic structure,
we have measured several Bragg intensities using the (100)/(010) and 
(110)/(001) orientations. The observations were corrected for the 
broadening in order to obtain integrated intensities.
The last column in table 1 gives the seven examined intensities relative to
the strongest one, (0.307 0.307 1). 
Since the propagation vector
is determined, only a few simple models may be applied.
As the propagation vector is not along the tetragonal axis,
sine-wave modulations with longitudinal and transverse
character are likely, 
whereas, a helimagnetic structure would imply some 
accidental degeneracy.
We have considered different models : with the magnetic moment along 
$c$, M1, with an in-plane moment 
perpendicular to the modulation, M2,  with in-plane moment 
parallel to [110], M3, and the less likely
helimagnetic structure, M4 \cite{models}.
The calculated intensities are given in 
table 1 relative to that of (0.307 0.307 1).
The intensity ratios are determined with good accuracy since 
the severe corrections for broadening and extinction
almost cancel out. The best description is obtained 
with M1, the sine-wave with the moments parallel $c$; models
M2 and M3 are ruled out. Also the helimagnetic model
yields a description worse than M1, but a minor in-plane contribution
of the ordered moment may not be excluded. Nevertheless, this analysis
shows that the main part of the ordered moment (possibly all of it)
is pointing along $c$.

\begin{table}
\begin{tabular}{c|c|c|c|c|c}
~~(hkl)-indices~~ & M1   & M2   & M3   & M4   &  observed \\
\hline
(0.307 0.307 1)   & 1    &  1   & 1    & 1    & 1       \\
(0.307 0.693 0)   & 0.61 & 0.05 & 1.10 & 0.28 & 0.51(4) \\
(0.307 0.307 3)   & 0.05 & 0.19 & 0.46 & 0.13 & 0.08(3) \\
(0.693 0.693 1)   & 0.23 & 0.17 & 0.05 & 0.20 & 0.27(5) \\
(0.693 0.693 3)   & 0.06 & 0.07 & 0.10 & 0.07 & 0.0(1)  \\
(1.307 0.307 0)   & 0.11 & 0.05 & 0.06 & 0.07 & 0.17(5) \\
(1.307 0.693 1)   & 0.07 & 0.01 & 0.14 & 0.06 & 0.07(2) \\
(0.307 0.307 5)   & 0.00 & 0.04 & 0.12 & 0.03 & 0.00(4) \\
\end{tabular}
\label{table1}
\caption{ Comparison of the measured 
integrated Bragg intensities (last column) to those 
calculated with the models M1-M4, see text.
All intensities are given  
relative to that of (0.307 0.307 1)  ($\lambda$=4.25\AA)
taking into account the form-factor
of Ru$^{1+}$.
The calculation assumes perfect averaging of domains and was performed using 
the FULLPROF-program \cite{fullprof}.
}
\end{table}

The estimation of the amplitude of the modulation is less straightforward,
since the corrections for the broadening and for extinction enter. 
From the integrated intensity of the magnetic (0.307 0.307 1) Bragg-peak
relative to that of (1 1 0) we obtain the ordered moment to 0.3(1)$\mu _B$;
the error takes account of the reliability of the corrections.
We emphasize that the ordered moment may not be 
below 0.1$\mu _B$, since this
would imply intensities a factor of 10 smaller.

The ordered moment is much larger than the 
ferromagnetic moment induced by a field of 1T at low temperature, 
0.005$\mu _B$ per Ru,
and it is still larger than the induced moment at higher temperature,
which was obtained by a Curie-Weiss
fit to the susceptibility \cite{minakata}. 
Therefore, one has to interpret the incommensurate order as arising 
directly from the intrinsic Fermi-surface nesting \cite{mazin,sidis}.
In the RPA treatment of itinerant magnetism the Stoner enhancement 
of the bare magnetic susceptibility, $\chi _0$,  through the interaction, 
$I$, is described by :

$$ \chi(q) = {{\chi _0(q)}\over {1-I\cdot\chi _0(q)}}~~~~(1). $$

In the pure compound we have found that the product 
$\alpha = I\cdot\chi _0(q)$ amounts
to 0.97 \cite{sidis}, yielding a susceptibility
enhancement at the nesting vector by a factor of 33.  
Ti-substitution appears to shift $\alpha$ towards the critical value
of one, which implies the divergence of the
susceptibility and the magnetic transition.

The magnetic excitation spectrum in \srton ~ has been studied 
by inelastic neutron scattering using the same crystal.
The constant energy scans
across {\bf Q$_i$} at energies above 3meV yield peaks 
at the nesting vector very similar to the observations in 
pure Sr$_2$RuO$_4$, see figure 3 \cite{sidis}.
These scans appear to sense an excitation continuum
similar to the paramagnetic phase ,
in particular there is no sizeable dependence of the peak 
width on the energy \cite{friedt}.
The correlation length of the excitations is comparable
to that found in the pure compound, $\sim$15\AA, but it is much shorter
than the correlation of the elastic order.
Magnetic order due to Fermi-surface nesting should open a partial gap
separating the high energy continuum from low energy spin-waves.
One of the very few well studied examples for such behavior is found in 
V$_{2-x}$O$_3$ \cite{bao,v2o3}.
A partial gap in \srton ~
agrees also with the increase in the resistivity \cite{minakata}.
The scans at 2.1\ meV indicate that the signal becomes
broadened and reduced in amplitude in the incommensurate ordered phase
in qualitative agreement with a gap and the spin-wave picture expected for 
low energies. 
These observations further support the interpretation, that the 
magnetic ordering results directly from the Fermi-surface nesting. 
A more quantitative analysis of the excitation spectrum, however,
will require larger crystals.

\begin{figure}[t]
\includegraphics*[width=0.55\columnwidth]{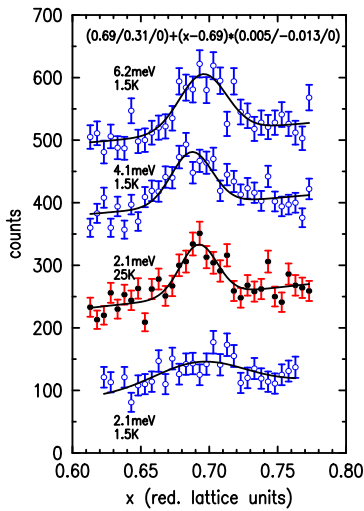}%
\caption{ Constant energy scans across the incommensurate
signal at {\bf Q$_i$}=(0.693 0.307 0), all scans were performed 
transverse to {\bf Q$_i$}. For the two scans at 4.1 and 6.2 meV performed
at 1.5 K the background was shifted by 150 counts.
For the scan at 2.1 meV and 1.5 K the  background was shifted by 
-150 counts.
Lines correspond to Gaussians and a sloped background.
} \label{fig:3}
\end{figure}

The spin-glass-like behavior at low temperature observed in ref. 
\cite{minakata} may result from the
disorder between Ru and non-magnetic Ti. 
On the one hand Ti substitution
drives the system towards magnetic ordering, and on the other hand, Ti
impurity acts as disorder centers, giving rise to finite correlations
lengths.
The average Ti-distance being of the order of the correlation length,
these vacancy moments may be aligned under an applied field
which may explain the observed weak ferromagnetic moments \cite{minakata}.
Furthermore, the rearrangement of the ordered clusters will
imply some relaxation processes.

In the context of the stripe scenario incommensurate magnetic order 
has gained interest also in the cuprates \cite{tranquada}. 
There also, the transition from quasi-elastic to elastic order
is sluggish.
Furthermore, the {\bf Q}-broadening 
of the elastic signal induced by  replacement of Cu by 
non-magnetic Zn in the cuprates \cite{cuprates} resembles our
observation in \srto . 
In the cuprates, however, the spin density wave is considered to 
be strongly coupled to a charge modulation, whereas 
it seems to be purely electronic in the ruthenate.

Another important impact of the magnetic ordering in \srto ~ concerns the 
orientation of the spins along $c$. Several groups have analyzed the problem
whether the incommensurate magnetic fluctuations may imply
spin-triplet superconductivity \cite{sigrist-n,japan1,japan2} 
which seems to be established in \srruo \cite{ishida}.
Two groups conclude that this would require some anisotropy of the
incommensurate fluctuations.
The required dominance of the out of plane component may be deduced 
from the NMR studies \cite{mukuda} which are, however, not restricted to 
the incommensurate wave-vector. The observation of the 
incommensurate ordering with the spins along $c$ in \srton ~
clearly indicates 
that the incommensurate fluctuations exhibit the required 
anisotropy at least qualitatively.

In conclusion, elastic neutron scattering on \srton ~ reveals 
incommensurate magnetic ordering. The large ordered moment
and the spin exictation spectrum
indicate that magnetic order is directly induced by the 
intrinsic Fermi-surface nesting instability. 
The pure compound, \srruo , is, hence, close to a quantum critical point
similar to cuprate or heavy fermion superconductors \cite{heavyf}.

We gratefully acknowledge discussions with P. Pfeuty.


\begin{thebibliography}{}
\bibitem{models}Models M1 and M3 may be coupled by symmetry.
\end{thebibliography}
\end{document}